\documentclass[12pt,unisize]{article}
\begin{document}
\title{GRAVITATIONAL ULTRARELATIVISTIC\\
SPIN-ORBIT INTERACTION\\
AND THE WEAK EQUIVALENCE PRINCIPLE}
\author{Roman Plyatsko
\thanks{
Electronic address: plyatsko@lms.lviv.ua}
\\
Pidstryhach Institute for Applied Problems \\ in Mechanics and
Mathematics,\\
National Academy of Sciences,\\
Naukova St. 3-b, Lviv, 79060, Ukraine}
\date{\vglue-2.5cm}
\maketitle
\begin{abstract}
It is shown that the gravitational
ultrarelativistic spin-orbit interaction violates
the weak equivalence principle in the traditional
sense. This fact is a direct consequence of the
Mathisson-Papapetrou equations in the frame of
reference comoving with a spinning test body. The
widely held assumption that the deviation of a spinning
test body from a geodesic trajectory is caused
by tidal forces is not correct.
\end{abstract}

\noindent PACS number(s):~04.20.Cv.~95.30.Sf

\section{Introduction}

The equivalence principle had attracted considerable attention in the
past decades \cite{1}, especially after the criticism of Fock \cite{2}
and Synge
\cite{3} concerning its sense and role in Einstein's gravitational theory.
Now this principle gives rise to renewed interest, particularly
in the context of new tests [4--9] (many papers have appeared
for the last years \cite{10}). In spite of the fact that in known books the
equivalence principle is presented as a cornerstone of the general relativity
theory, and a principle of prime importance (see, for example, \cite{11}),
the doubts generated by Fock and Synge have not disappeared completely.

The purpose of this paper is to show that there is an objective reason
for the revision of the physical content and meaning of the equivalence
principle in general relativity. This reason is based on results of a
more careful analysis of the spinning test body deviation
from the geodesic motion in the gravitational field.

As regards the influence of spin (inner rotation) of a body on its motion in
the gravitational field, the prevailing opinion is presented
particularly well in Ref.\cite{11}, exercise 40.8. According to this opinion
the deviation of a spinning test body from the geodesic trajectory is
caused by tidal forces connected with the spacetime curvature. The
Mathisson-Papapetrou equations (MPE) \cite{12,13} are singled out
as providing information on the behavior of a spinning test body in
a gravitational field. However, the appeal to these equations in
\cite{11} is not sufficiently justified. The arguments are patently
insufficient for assertion that the
interaction between the spin of a test body and the spacetime
curvature is reduced to the influence of tidal forces. Therefore, here we
shall direct our attention to the MPE. Although these
equations were extensively investigated in the 1960s and 1970s a number of
problems remain. Particularly regarding the influence of spin on
the world line of a test body \cite{14}. Certainly, if one starts from
general considerations, relying on the equivalence principle, then it is
\mbox{a priori} clear that spin can only slightly deform the world
line of a test body, as compared to the corresponding geodesic line.
But there is another way, namely, to forget for the time being about
the equivalence principle, and try to discover facts that follow
directly from the MPE, without \mbox{{\bf a priori}} restrictions.

Here we shall consider the consequences of the interaction between the spin of
a test body and the curvature of a Schwarzschild's field and compare these
with properties of tidal forces in this field. In particular, it is known
(see, e.g., equations (31.6) and (32.24a) in
\cite{11}) that for radial motions in this field tidal forces
are determined by the nonzero components of Riemann's
tensor in the comoving frame of reference:
\begin{equation}\label{1}
\parbox{4cm}{
\begin{eqnarray*}
R_{\hat{\tau}\hat{\rho}\hat{\tau}\hat{\rho}}&=&-\frac{2m}{r^3},\\
R_{\hat{\theta}\hat{\phi}\hat{\theta}\hat{\phi}}&=&\frac {2m}{r^3},
\end{eqnarray*}}
\qquad
\parbox{6cm}{
\begin{eqnarray*}
R_{\hat{\tau}\hat{\theta}\hat{\tau}\hat{\theta}}&=&
R_{\hat{\tau}\hat{\phi}\hat{\tau}\hat{\phi}}=\frac m{r^3},\\
R_{\hat\rho\hat\theta\hat\rho\hat\theta}&=&
R_{\hat\rho\hat\phi\hat\rho\hat\phi}=-\frac m{r^3}
\end{eqnarray*}}
\end{equation}
(in notation of \cite{11} for local indices; $m$ is the Schwarzschild
mass). Namely, the comoving frame of reference was used for
the analysis of tidal forces in \cite{11}, sections 31.2 and 32.6.
For the direct and correct comparison of tidal forces with the
forces caused by spin-curvature interaction it is expedient to
consider the consequences of MPE in the comoving frame of reference
also.

We shall investigate the spin-curvature interaction not only for
radial motions because the weak equivalence principle is formulated
for general motions (see, e.g., section 2.3 of \cite{14}).
For example, it is applied to the usual falling lift, as well as to the
satellites orbiting the Earth.

\section{The MPE in a comoving frame of reference}

The traditional form of the MPE is \cite{12,13}
\begin{equation}\label{2}
\frac D {ds} \left(Mu^\lambda + u_\mu\frac {DS^{\lambda\mu}} {ds}\right)=
-\frac {1} {2} u^\pi S^{\rho\sigma} R^{\lambda}_{\pi\rho\sigma},
\end{equation}
\begin{equation}\label{3}
\frac {DS^{\mu\nu}} {ds} + u^\mu u_\sigma \frac {DS^{\nu\sigma}} {ds} -
u^\nu u_\sigma \frac {DS^{\mu\sigma}} {ds} = 0,
\end{equation}
where $u^\lambda$ is the 4-velocity of a spinning test body (STB),
$S^{\mu\nu}$ is the tensor of spin, $M$ and $D/ds$ are, respectively,
the mass and the covariant derivative. For the description of the test
body center of mass, Eqs.~(\ref{2}),~(\ref{3}) are
often supplemented by the relation~\cite{12,15,16}
\begin{equation}\label{4}
S^{\mu\nu} u_\nu = 0.
\end{equation}
The correct definition of the center of mass for a STB is a subject of
discussions \cite[17--22]{14}. Below we shall
investigate the MPE in the approximation linear in spin (in
accordance with the linear consideration of tidal forces in \cite{11}),
when supplementary condition~(\ref{4}) coincides with the
alternative condition proposed by Tulczyjew \cite{19} and Dixon
\cite{20,21}. Here we shall not consider discussions of \cite[17--22]{14}.

Besides $S^{\mu\nu}$, the 4-vector of spin $s_\mu$ is also used in the
literature where by definition~\cite{16}
$$
s_\mu = {1\over2} \sqrt{-g} \epsilon_{\mu\nu\rho\sigma} u^\nu S^{\rho\sigma}
$$
($g$ is the determinant of the metric tensor).

For transformations of Eqs.~(\ref{2}),~(\ref{3}) we use the
known relations for the orthogonal tetrads $\lambda_\mu^{(\alpha)}$,
\begin{equation}\label{5}
g_{\mu\nu} =
\lambda_\mu^{(\alpha)}\lambda_\nu^{(\beta)}\eta_{(\alpha)(\beta)},\qquad
\lambda_\mu^{(\alpha)}\lambda_{(\beta)}^\mu = \delta_{(\beta)}^{(\alpha)},
\end{equation}
($\eta_{(\alpha)(\beta)} $ is the Minkowski tensor) and the conditions
for comoving tetrads \cite{16}
\begin{equation}\label{6}
dx^{(i)} = \lambda_\mu^{(i)}dx^\mu = 0,\qquad
 dx^{(4)} = \lambda_\mu^{(4)}dx^\mu = ds,
\qquad \lambda_{(4)}^\nu = u^\nu
\end{equation}
(here and in the following, indices of the tetrad are placed in the
parentheses; latin indices run $1,\,2,\,3$ and greek indices $1,\,2,\,3,\,4$).
For convenience, we choose the first local coordinate axis as orientated
along the spin, then
\begin{equation}\label{7}
s_{(1)} \neq 0,\qquad
s_{(2)} = 0,\qquad
s_{(3)}=0,\qquad
s_{(4)}=0
\end{equation}
and $|s_{(1)}|=S_{0}=\mbox{const}$. is the
value of the spin of a test body as measured by the comoving observer
\cite{16}.

From Eq.~(\ref{3}), taking into account~(\ref{4})--(\ref{7}), we
obtain $\gamma_{(i)(k)(4)}=0$, the known condition for the
Fermi-Walker transport, where $\gamma_{(\alpha)(\beta)(\delta)}$ are
Ricci's coefficients of rotation~\cite{23}.

From~(\ref{2}), after corresponding calculations, we find
\begin{equation}\label{8}
\gamma_{(i)(4)(4)}+{s_{(1)}\over M}R_{(i)(4)(2)(3)}=0.
\end{equation}

It is important that Ricci's coefficients of rotation $\gamma_{(i)(4)(4)}$
have a direct physical meaning, namely, as the components $a_{(i)}$ of the
3-acceleration of a STB relative to geodesic free fall as measured by the
comoving observer (one can see this fact from the equation of geodesic
world lines in a vector tetrad space
$\ddot x^{(i)}+\gamma^{(i)}_{(\alpha)(\beta)}\dot x^{(\alpha)}
\dot x^{(\beta)}=0$). Therefore, by Eq.~(\ref{8}) we have
\begin{equation}\label{9}
a_{(i)} =-{s_{(1)}\over M} R_{(i)(4)(2)(3)}.
\end{equation}

\section{Spin-curvature interaction in \\ Schwarzschild's field}

For simplicity, we shall restrict ourself to the case of equatorial
motions of a STB in Schwarzschild's field when spin is orthogonal to the
motion plane. Using the components of the metric tensor in the standard
coordinates $x^1=r$, $x^2=\theta=\pi/2$, $x^3=\varphi$, $x^4=t$ and
relations~(\ref{5}),~(\ref{6}) it is easy to find the nonzero components
of the comoving tetrads:
$$
\lambda^2_{~(1)}=\sqrt{-g^{22}}, \quad
\lambda^1_{~(2)}=u^1u^4\sqrt{\frac{g_{44}}{u_4u^4-1}}, \quad
\lambda^3_{~(2)}=u^3u^4\sqrt{\frac{g_{44}}{u_4u^4-1}},
$$
$$
\lambda^4_{~(2)}=\sqrt{\frac{u_4u^4-1}{g_{44}}}, \quad
\lambda^1_{~(3)}=u^3\sqrt{\frac{g^{11}g_{33}}{u_4u^4-1}},
$$
\begin{equation}\label{10}
\lambda^3_{~(3)}=-u^1\sqrt{\frac{g^{33}g_{11}}{u_4u^4-1}}, \quad
\lambda^1_{~(4)}=u^1, \quad \lambda^3_{~(4)}=u^3, \quad
\lambda^4_{~(4)}=u^4.
\end{equation}
The nonzero components of the Riemann tensor in the standard
coordinates for $\theta=\pi/2$ are given by
$$
R_{1212}=R_{1313}=\frac mr\left(1-\frac{2m}{r}\right)^{-1}, \quad
R_{2323}=-2mr,
$$
\begin{equation}\label{11}
R_{1414}=\frac{2m}{r^3}, \quad R_{2424}=R_{3434}=-\frac mr
\left(1-\frac{2m}{r}\right)
\end{equation}
(signature ---,---,---,+).
For calculation of the local components of the Riemann tensor
we use the general relation
\begin{equation}\label{12}
R_{(\alpha)(\beta)(\gamma)(\delta)}=\lambda^\mu_{(\alpha)}\lambda^\nu_{
(\beta)}\lambda^\rho_{(\gamma)}\lambda^\sigma_{(\delta)}R_{\mu\nu\rho\sigma}.
\end{equation}
Inserting Eqs.~(\ref{10}) and~(\ref{11}) into Eq.~(\ref{12})
we obtain the expressions for the components $R_{(i)(4)(2)(3)}$:
$$
R_{(1)(4)(2)(3)}=0,\qquad
R_{(2)(4)(2)(3)}=-{3mu^1u^3\over{r^2\sqrt{u_4u^4-1}}}\,\left (1-
{2m\over{r}}\right )^{-1/2},
$$
\begin{equation}\label{13}
R_{(3)(4)(2)(3)}=-\frac{3mu^3u^3u^4}{r\sqrt{u_4u^4-1}}
\left(1-{2m\over{r}}\right)^{1/2}.
\end{equation}
By Eq.~(\ref{9}) these local components of the Riemann tensor determine
the force of the spin-curvature interaction from the point of view of a
comoving observer.

Now we can compare components~(\ref{13}) with local (comoving) components
of the Riemann tensor~(\ref{1}) which
determine tidal forces for radial motions.
It is easy to see that for radial motions, when $u^3=0$, all components
(\ref{13}) are equal 0. That is, in this case the spin-curvature interaction
{\bf does not} deviate the motion of a spinning test body from the geodesic
radial motion. (This fact is also known from the partial solution of
Eqs.~(\ref{2})--(\ref{4})
in the Schwarzschild field). At the same time, all components (\ref{1})
and tidal forces for radial motions {\bf are not} equal 0. [The correspondence
between the local indices in Eq.~(\ref{1}) and the notation in (\ref{13}) for
radial motions is given by $\hat {\rho} \to (2)$, $\hat {\theta} \to (1)$,
$\hat {\phi} \to (3)$, $\hat {\tau} \to (4)$. Then in accordance with (\ref{1})
such components of the Riemann tensor in our notation are not equal~0:
\begin{equation}\label{14}
R_{(i)(4)(i)(4)}, \qquad R_{(i)(j)(i)(j)}
\end{equation}
($i\ne j$).
The expressions for these components follow directly from relation (\ref{10})
(at $u^3=0$), (\ref{11}), (\ref{12}) and one can check that they coincide with
the corresponding right-hand sides of Eq.~(\ref{1})].

We emphasize that each component of~(\ref{13}) has only one time local
index, whereas such components are absent among those of Eq.~(\ref{14}).
This fact is very important because the components of the Riemann tensor
with one time index correspond to the "gravitomagnetic" components of
the gravitational field \cite{24, 25}. The components of the Riemann
tensor with two time indices correspond to the
"gravitoelectric" components of the gravitational field. According to
the unnumbered equation preceding Eq.~(32.24b) of \cite{11} namely the
"gravitoelectric" components cause tidal forces.
(The deep analogy between the "gravitomagnetic moment" of a spinning test
body in general relativity and the magnetic moment in electromagnetism was
studied in \cite{22}).

So, we cannot consider tidal forces as the reason for the
STB deviation from the geodesic motion.

We can indicate two other arguments in support of the conclusion that the MPE
do not contain tidal forces. For example, let us suppose that
the tidal forces are present in the MPE. Then these forces cannot
disappear if we make the spin equal to zero
(more exactly, the angular velocityof the inner rotation)
in the MPE,
because the tidal forces are connected with the dimension of a test
body and its nonrotating state does not remove these forces. However,
if one puts $S^{\mu\nu}=0$, Eqs. (2), (3) pass to the geodesic eqs.
and do not to the eqs. of the geodesic deviation. The geodesic
eqs. do not contain tidal forces (in contrast to the geodesic
deviation eqs., which do) and therefore the assumption that tidal forces
are present in the MPE is not correct.

It is necessary to remember that the tidal forces will be taken into
account if we consider two close world lines (see Ref.[26], Chap.~6,
Sec.~10, where the clear procedure for derivation of geodesic
deviation eqs. is given). However, the MPE, as well as the geodesic
equations, describe {\bf only one} world line. Therefore, we can point out
the place in the procedure of the MPE derivation where the tidal forces
were neglected: when the world tube of a test body was replaced with
{\bf only one} world line (see, e.g., Ref.[13], page 250).

The second additional argument, which refutes the assumption that presence
of tidal forces is the reason of the STB deviation from the geodesic
motion, is connected with the known fact that the MPE are the classical
limit of the general relativistic Dirac equation. In a number of
publications it is shown that the right-hand side of Eq. (2) describes
the interaction of a quantum electron with a gravitational field \cite{27}.
Obviously, one cannot speak about tidal forces in the Dirac equation.

We stress that the authors of Ref.[11] do not provide a proof of the
statement that the tidal forces are present in the MPE; it is an
assumption (hypothesis) only. The appearance of the Riemann tensor in
the right-hand side of Eq.~(2) is not a sufficient argument for this
statement, because this tensor has a number of physically different
components and only part of them is connected with tidal forces.
Our rigorous consideration, presented above, gives a direct proof of
the conclusion that the right-hand side of Eq.~(9) does not contain
any components of the Riemann tensor that are connected with tidal forces.

If tidal forces are not the reason for the deviation of a
STB from the geodesic
trajectory, then what is the reason? Considering~(\ref{9}) and~(\ref{13})
it is easy to answer this question. Indeed, for the value of
3-acceleration $\vert \vec a\vert $ of a STB relative to geodesic free
fall, where
$$
\vert \vec a\vert =
\sqrt{a^2_{(1)}+a^2_{(2)}+a^2_{(3)}},
$$
using~(\ref{9}),~(\ref{13})
we find for all cases of equatorial motions (not only for the
circular orbits)
\begin{equation}\label{15}
\vert \vec a\vert =
{m\over{r^2}}{3S_0\vert u_\perp \vert\over{Mr}} \sqrt{1+u^2_\perp },
\end{equation}
where $u_\perp = r\dot \varphi$ is the tangential
component of the test body 4-velocity. Even though Eq.~(\ref{13}) contains
the radial velocity, when we calculate $|\vec a|$, the terms with
$u^1$ cancel out due to the relation $u_\mu u^\mu=1$.
(By~(\ref{9}),~(\ref{13})
one can check that vector $\vec{a} $ is orientated towards the source
of the Schwarz\-schild field).  In accordance with Eq.~(\ref{15}) $\vert
\vec a\vert$ is nonzero only if $u_\perp \ne 0$. When the velocity of a
STB is much less than the velocity of light, i.e. when $\vert u_\perp
\vert\ll 1$, Eq.~(\ref{15}) corresponds to the expression~(44) from
paper~\cite{22}, where the spin-spin and spin-orbit gravitational
interactions were investigated in the lowest approximation in the
velocity of a STB. One may consider Eq.~(\ref{15}) as the
generalization of Eq.~(44) from~\cite{22} for any velocities of a STB.
This generalization is not trivial and contains significant new
information on gravitational spin-orbit interaction. Namely, when
$\vert u_\perp \vert\ll~1$ after Eq.~(\ref{15}) and the condition for a
STB~\cite{22}
\begin{equation}\label{16}
S_0/Mr\ll 1
\end{equation}
we have $\vert \vec
a\vert \ll m/r^2$, where $m/r^2$ is the Newtonian value of the free
fall acceleration. In this case, if the dimension of a STB (and its
value $S_0$) are sufficiently small, the $\vert \vec a\vert$ is
negligible and we can say that gravitational spin-orbit interaction
obeys the weak equivalence principle. However, another situation in
principle exists in the ultrarelativistic region of velocities, when
$\vert u_\perp \vert \gg 1$. Then, according to~(\ref{15}), for any small
$S_0/Mr$ we can indicate such sufficiently large value $\vert u_\perp
\vert$ for which the value of acceleration of a STB measured by the
comoving observer will not be negligible. Therefore, the
ultrarelativistic gravitational spin-orbit interaction violates the
weak equivalence principle in the traditional sense. Here we accent that
the usual formulation of the weak equivalence principle is not restricted
to the special case of the radial fall. This principle is applied to any
motion in any gravitational field. The partial result that for $u_{\perp}=0$
from Eq.~(\ref{15}) we have $|\vec {a}|=0$ cannot remove the necessity of
reinterpretation of the weak equivalence principle. Expression Eq.~(\ref{15})
demonstrates the limit of validity of this principle in the traditional
formulation.

Thus, in accordance with~(\ref{15}) and~(\ref{16}), the above two limiting
processes are essentially different in their physical consequences.
\begin{enumerate}
\item
The dimension of a test body and its spin
tend to $0$ whereas the body velocity is limited from above and
fixed (but as close as one likes to the velocity of light);
\item
The dimension and spin of a test body is as small as one likes, but
fixed, but larger and larger velocity is given to a body. In the
first case, the motion of a STB tends to the geodesic motion, while in the
second case a STB is moving away from it more and more. If in the
first case the weak equivalence principle is fulfilled, then in the
second case it is obviously violated. In regard to the corresponding
control experiment one cannot assert that it does not depend on the
velocity of the free falling device of Ref.~\cite{14}, Sec.~ 2.3.
\end{enumerate}

One can rewrite Eq.~(\ref{15}) in the form
\begin{equation}\label{17}
|\vec a|=\frac{m}{r^2}\frac{3S_0|L|}{M^2r^2}\sqrt{1+u^2_\perp },
\end{equation}
where $L$ is the orbital momentum of a test body. In Ref.\cite{28}
R.~Micoulaut has investigated the connection between the orbital
momentum and the velocity of a spinning test body for equatorial
motions in a Schwarzschild field. From the results of \cite{28}
it is clear that $L$ is arbitrarily large for any solution of the MPE
with the sufficiently large initial value of the tangential component
of the test body 4--velocity. So, the MPE admit the motions with
arbitrarily large $L$. On the whole, for these motions $r$ is not
const. We emphasize that Eqs.~(\ref{15}) and (\ref{17}) are valid for
all cases of equatorial motions of a STB in a Schwarzschild field,
including $r\ne const.$

Naturally, under our earthly conditions we do not have the possibility to
launch a macroscopic ultrarelativistic STB with the comoving observer for
the registration of acceleration~(\ref{15}). For the
ultrarelativistic elementary particles
we cannot realize a comoving ultrarelativistic falling laboratory.
[It follows from Eq.~(\ref{15}) that the electron flying near the
surface of the Earth feels the acceleration equal to $9.8\,ms^{-2}$, i.e. to
the Newtonian acceleration for the Earth, if the velocity of an electron
corresponds to the energy of its free motion $\approx 10^{15}\,eV$.
One may obtain this value substituting $S_0=h/2$ in Eq.~(\ref{15})].
Nevertheless, in principle, the possibility exists for
a nondirect examination of Eq.~(\ref{15}) by an earthly observer,
because for such an observer an ultrarelativistic electron or proton
must change its electromagnetic radiation due to the additional
nongeodesic acceleration connected with Eq.~(\ref{15}).

\section{Conclusions}

The importance of the expression~(\ref{15}) is not
restricted to the assertion that the gravitational spin-orbit
interaction in the ultrarelativistic limit indicates the necessity for a
more careful wording of the weak equivalence principle. Namely,
according to Eq.~(\ref{15}) we cannot treat this principle {\bf at all} as
one that excludes the gravitational field as a force field and which, at the
same time, interprets all gravitational actions as inertial actions. Quite on
the contrary, we see a closer analogy between gravitation and
electromagnetism. In electromagnetism for the
demonstration of the presence of a magnetic field it is necessary to have,
for example, a magnetic needle. Similarly, in the case of the gravitational
field, the presence of a spinning test body, serving as a specific
"gravitomagnetic needle", allows us to show that in a free falling frame of
reference the gravitational field
{\bf does not disappear}. Only a test body without spin does not feel this
field.

Thus, the traditional interpretation of the weak equivalence principle must
be revised {\bf in principle}. In this connection, we remember the considered
judgement of Fock that the kinematic interpretation of gravity plays a
heuristic role only. According to Fock, the true logical foundation of
Einstein's gravitational
theory is not the equivalence principle but other two ideas, namely, the
idea of spacetime unification in the united 4-dimensional chronogeometrical
manifold with an indefinite metric and giving up the "rigidity" of the
metric, which allowed to unite it with the phenomenon of gravity.

It is necessary to analyze the gravitational phenomena for which the weak
equivalence principle is violated as a consequence of the gravitational
ultrarelativistic spin-orbit interaction. In all probability, the necessity
will arise for a modification of the traditional picture of the
gravitational collapse because it is based on the analysis of the geodesic
world lines only.

\bigskip
\centerline{\bf Acknowledgment}

The author would like to thank Prof. O.M.Bilaniuk for his support and
encouragement and for a helpful reading of this manuscript.

\end{document}